\documentclass[nofootinbib,twocolumn,aps,prd,amsmath]{revtex4-2} 
\usepackage{amsmath}
\usepackage{txfonts}
\usepackage{amssymb}
\usepackage{mathrsfs}
\usepackage{graphicx}
\usepackage{epsfig}
\usepackage{dcolumn}
\usepackage{bm}
\usepackage{ulem}
\usepackage{color}
\newcommand{\fw}[1]{{\color[rgb]{1.00,0.00,0.00} #1}}

\begin{document}

\title{Strange stars admixed with mirror dark matter: confronting
  observations of XTE J1814-338}


\author{Shu-Hua Yang$^{1}$}\email{ysh@ccnu.edu.cn}
\author{Chun-Mei Pi$^{2}$$^,$$^3$}
\author{Fridolin Weber$^4$$^,$$^5$}\email{fweber@ucsd.edu}

\affiliation{$^1$Institute of Astrophysics, Central China Normal
  University, Wuhan 430079, China\\$^2$School of Physics and
  Mechanical \& Electrical Engineering, Hubei University of Education,
  Wuhan 430205, China \\$^3$Research Center for Astronomy, Hubei University of Education, Wuhan 430205, China\\$^4$Department of Physics, San Diego State
  University, San Diego, CA 92182, USA \\$^5$ Department of Physics, University of
  California at San Diego, La Jolla, CA 92093, USA
}

\date{October 2024}

\begin{abstract}
  In this paper, we explore a novel framework for explaining the mass
  and radius relationships of observed neutron stars by considering
  strange stars (SSs) admixed with mirror dark matter (MDM). We develop
  a theoretical model that incorporates non-commutative algebra to
  describe the interactions between ordinary strange quark matter
  (SQM) and MDM, which are predicted to form compact objects that
  could explain recent astrophysical data, including observations of
  PSR J0740+6620, PSR J0030+0451, PSR J0437-4715, and the central
  compact object in HESS J1731-347. Notably, we demonstrate that the
  exotic mass-radius measurement of XTE J1814-338 can be explained by
  the presence of a mirror SS with an ordinary SQM core. In contrast
  to other explanations based on boson stars, our SS+MDM model offers
  a natural explanation for this system. We provide detailed
  mass-radius comparisons with observational data and discuss future
  observations that could test the predictions of our model, offering
  new insights into neutron star structure and the role of dark matter
  in compact objects.
\end{abstract}

\maketitle

\section{INTRODUCTION} \label{S:intro}

\fw{}





Neutron stars (NSs) have long been a subject of intense study due to
their extreme physical properties and their role as cosmic
laboratories for dense matter physics. Traditionally, NSs are modeled
as compact objects composed of nuclear matter. However, recent
observations have challenged these conventional models, including
measurements of NS masses, radii, and tidal deformabilities derived
from gravitational waves, X-ray timing, and thermonuclear burst
oscillations. Notably, NICER observations of PSR J0740+6620 and PSR
J0030+0451, along with the exotic source XTE J1814-338, suggest that
NSs may possess more exotic compositions than previously thought.
In particular, the compact stellar object XTE J1814-338 defies
conventional explanations. This object exhibits an unusually low
radius $(\sim 7$~km) and mass ($\sim 1.2\, M_\odot$).
 These
observations call for new theoretical models that extend beyond
standard nuclear matter and explore the possibility of exotic matter
within NSs.

One such possibility is that NSs could actually be strange stars (SSs)
\citep{Alcock1986,Bombaci1997,AlcockOlinto1988,Madsen1999,Weber2005,Zhang2024},
which are made of strange quark matter (SQM). SQM, which consists of
up ($u$), down ($d$), and strange ($s$) quarks, was proposed by Witten
\cite{Witten1984}, Farhi and Jaffe \cite{Farhi1984}, and others as a
potential ground state of matter, even more stable than nuclear
matter. SSs could explain the existence of NSs with unusually high
masses and small radii, such as PSR J0740+6620, without invoking
exotic particles beyond the Standard Model. In this scenario, the
entire star is composed of deconfined quarks, unlike conventional NSs
which are primarily made of neutrons.

However, SSs alone may not fully account for recent
observations. Mirror dark matter (MDM) has emerged as a promising
extension to the SS model. MDM, first proposed by Foot et al. \cite{Foot1991} and
later expanded by Berezhiani et al. \cite{Berezhiani2021}, is a stable, self-interacting dark matter candidate that interacts with ordinary
matter only through gravity. The concept of a dark matter halo
surrounding an SS could help to explain the peculiar mass and radius
observations of XTE J1814-338 \cite{Kini2024}. This star exhibits an unusually small
radius ($R \approx 7$ km) and relatively low mass ($M \approx 1.2
M_{\odot}$), which are difficult to reconcile with standard NS or even SS models. We propose that this compact object could
be a mirror strange star (MSS), with an SQM core
surrounded by an MDM halo. This dual-component model may explain both
the high compactness of XTE J1814-338 and the more typical mass-radius
relationships observed in other NSs and SSs.

Liu et al. \cite{Liu2024b} explored the impact of dark matter halos on the
pulse profiles of X-ray pulsars, showing that dark matter can
significantly modify pulse shapes and peak fluxes. This complements
our findings by illustrating how dark matter, whether mirror or
bosonic, could influence the observable properties of compact objects
like SSs. The presence of dark matter halos or admixtures,
as demonstrated in their study, could provide additional observational
evidence for dark matter in astrophysical systems.

Recently, Pitz and Schaffner-Bielich \cite{Pitz2024} proposed that XTE J1814-338 might be a boson star with a nuclear matter core, a model that shares some similarities with the MSS model. However, the boson star model requires the introduction of new scalar particles beyond the Standard Model, which complicates its physical interpretation. Our approach, by contrast, sticks to known particles and symmetries, making it a more conservative extension of the SS hypothesis.

Recent gravitational wave observations from events like GW170817 and
GW190814 have provided new constraints on the
equation of state (EOS) of NSs. These observations place upper limits on the
tidal deformability and radius of NSs, favoring stiff EOSs that are
more consistent with SQM or dark matter-admixed stars
\citep{Abbott2017,Abbott2018,Dittmann2024}. MDM admixed SSs offer a way to
reconcile these constraints with the observed compactness of sources
like XTE J1814-338.

In this paper, we aim to build on these ideas by exploring the
mass-radius relationships of SSs admixed with MDM, comparing them with recent observations. In particular, we
focus on explaining the properties of XTE J1814-338 and other peculiar
NS candidates that defy explanation through conventional
models. Our study expands upon previous work on SSs
\citep{Weber2005,Kuerban2020,Geng2021} and MDM \citep{Berezhiani2021, Yang2021b}, providing a unified
framework for understanding compact objects with extreme densities and
exotic compositions.

This paper is structured as follows. In Sec. \ref{EOS}, we present the
theoretical framework for SSs admixed with MDM, focusing on the EOS. Section
\ref{results} details the results, analyzing mass-radius relations and
parameter constraints. In Sec. \ref{sec:testing}, we discuss
observational implications and propose future tests for the model
using gravitational waves and X-ray timing. A summary is provided in
Sec. \ref{Summary}.

\section{Equation of State of Strange Quark Matter and Mirror
  Dark Matter}\label{EOS}

The equation of state (EOS) determines the relationship between
pressure and energy density, serving as the foundation for modeling
the properties of compact stars, such as their mass-radius
relationship, stability, and possible phase transitions.  In this
study, we employ a modified bag model for SQM, which accounts for key
properties such as charge neutrality and chemical equilibrium. To
incorporate MDM, we assume a symmetric EOS for mirror quark matter,
reflecting the same thermodynamic principles as ordinary SQM. This
unified EOS framework allows us to analyze the effects of MDM on the
mass-radius relationship and test its alignment with observational
constraints.

\subsection{Strange Quark Matter}

SQM is hypothesized to consist of roughly equal
numbers of up ($u$), down ($d$), and strange ($s$) quarks, along with
a small admixture of electrons to maintain charge neutrality. This
composition is predicted by the strange matter hypothesis, originally
proposed by Itoh \cite{Itoh1970} and further developed in seminal works
such as \cite{Bodmer1971,Witten1984,Farhi1984}. According to this hypothesis, SQM may be more
stable than ordinary nuclear matter, implying that compact stars could
exist as SSs rather than NSs.

For the EOS of SQM, we employ the modified bag model, which has been
extensively used in studies of SSs
\citep{Farhi1984,Alcock1986,Haensel1986,Weber2005}. In this model, the
quarks are considered to be massless (for $u$ and $d$ quarks), while
the $s$ quark has a finite mass ($m_s=93$ MeV \cite{Navas2024}). First-order perturbative corrections to the strong
interaction coupling constant $\alpha_{S}$ are included to account for
interactions among quarks. This model also incorporates the bag
constant ($B$), which represents the vacuum pressure that confines
quarks inside hadrons or within the SSs.

The number density for each species (quarks and electrons) is given by:
\[
n_{i}=-\frac{\partial\Omega_{i}}{\partial\mu_{i}},
\]
where $i = u, d, s, e$; $\Omega_{i}$ are the thermodynamic potentials 
for up, down, strange quarks, and electrons (the formalism of $\Omega_{i}$ can be found in Refs. \cite{Alcock1986,Yang2020}), and $\mu_i$ are the respective chemical
potentials. Chemical
equilibrium is maintained via weak interaction processes, resulting in
the relations:
\[
\mu_{d}=\mu_{s}, \quad \mu_{s}=\mu_{u}+\mu_{e}.
\]
Charge neutrality is enforced by the condition:
\[
\frac{2}{3}n_{u}-\frac{1}{3}n_{d}-\frac{1}{3}n_{s}-n_{e}=0.
\]
The energy density and pressure of SQM are then given by:
\[
\epsilon_{Q} = \sum_{i=u,d,s,e} (\Omega_{i} + \mu_{i} n_{i}) + B,
\]
\[
p_{Q} = -\sum_{i=u,d,s,e} \Omega_{i} - B,
\]
where $B$ is the bag constant, typically chosen to be within the range
of $B^{1/4} = 130 - 160$ MeV, depending on the specific model and
observational constraints.

The modified bag model, while effective in describing SQM, has certain limitations, such as the omission of higher-order
quantum corrections and possible phase transitions at extreme
densities. However, it remains one of the most widely used models due
to its simplicity and ability to reproduce key features of compact
stars in high-density regimes \citep{Weber2005,Weissenborn2011,Zhou2018,Alford2005,Bhattacharyya2016,Miao2021,Zhang2021,Oikonomou2023,Wang2024}.

\subsection{Mirror Dark Matter}

MDM is a candidate for the stable,
self-interacting dark matter that could coexist with ordinary
matter. This concept arises from parity-symmetric extensions of the
Standard Model, first proposed by Lee and Yang \cite{Lee1956} and Kobzarev et al.
\cite{Kobzarev1966}, and later expanded by others \cite{Blinnikov1982,Blinnikov1983,Khlopov1991,Foot1991,Mohapatra1997,Mohapatra2002}. In these
models, every Standard Model particle has a mirror counterpart, with
the key distinction that mirror particles interact via right-handed
interactions, while ordinary particles interact via left-handed
interactions. These mirror particles form a hidden, parallel universe,
with MDM potentially interacting only gravitationally with ordinary
matter.

The theoretical framework of MDM is reviewed in detail in Refs. \cite{Foot2004,Foot2014,Berezhiani2004,Berezhiani2018}, among others. In
this study, we assume the simplest case in which the microphysics of
MDM mirrors that of ordinary matter, implying that MDM follows the
same EOS as SQM. This means that the mirror SQM
(mirror up ($u'$), down ($d'$), and strange ($s'$) quarks) is governed
by the same thermodynamic principles as ordinary SQM.

In the minimal parity-symmetric extension of the Standard Model, the
gauge group is doubled, $G \otimes G$, where $G$ represents the
Standard Model gauge group. This symmetry ensures that the
interactions within the mirror sector are identical to those in the
ordinary sector, except for the handedness of the
interactions. Consequently, the EOS for the mirror SQM is identical to that of
SQM.

Recent studies, such as those by Liu et al. \cite{Liu2024a} and Mariani et al. \cite{Mariani2024}, have demonstrated the critical role of the dark matter self-interaction cross-section constraint ($\sigma/m$) in determining the observable properties of compact stars. This constraint, derived from galaxy cluster observations, provides a range ($\sigma/m \sim 0.1-10 \, \mathrm{cm^2/g}$) \cite{Tulin2018} that aligns with both astrophysical data and theoretical models. 

For MDM, the cross-section constraint derived from galaxy clusters observations might be related to the QCD scale of it ($\Lambda'$). Mohapatra et al. \cite{Mohapatra2002} found that for scenarios with appreciable mirror symmetry breaking, e.g., the QCD scale of the MDM ($\Lambda'$) is 30 times the ordinary matter ($\Lambda$), the mirror hydrogen atoms could satisfy the cross-section constraint derived from galaxy clusters observations. In this case, the maximum mass and the radius of the pure MSSs will be smaller than ordinary SSs (just like the case of mirror NSs  \cite{Hippert2022}), and the structure of the MDM admixed SSs will also be changed significantly (like MDM admixed NSs \cite{Hippert2023}).

However, Mohapatra and Nussinov \cite{Mohapatra2018} found that to fulfill the observation of $n-n'$($n'$ is the mirror neutron) oscillations, a very precise mirror symmetry is required (same result is found by Berezhiani \cite{Berezhiani2019}), which leads the result that the cross-section of mirror hydrogen atoms is much higher than the upper bound from galaxy clusters observations. Nevertheless, Mohapatra and Nussinov \cite{Mohapatra2018} proposed that if most of the MDM in galaxies and galaxy clusters reside in collisonless stars and not in gas, the cross-section constraint could be satisfied for mirror scenarios with precise mirror symmetry.

Thus, considering the current status about the study of the cross-section constraint to MDM, we will use the symmetric model ($\Lambda'=\Lambda$, where the EOS for the mirror SQM is the same as that of
ordinary SQM) following Mohapatra and Nussinov \cite{Mohapatra2018} and Berezhiani \cite{Berezhiani2019} in this paper.

\subsection{Interaction between SQM and MDM}

In our model, SQM and MDM interact solely through gravitational forces. Although the neutron–mirror neutron mixing has been widely studied \cite{Berezhiani2006,Berezhiani2009,Goldman2019,McKeen2021,Goldman2022}, the nature of direct interaction between quarks and mirror quarks remains speculative and unstudied. However, if
quark-mirror quark interactions exist, they are expected to be weak
enough to be neglected in our study, as suggested by Berezhiani et al. \cite{Berezhiani2021}. Therefore, we model the star as a
two-component system, where the gravitational coupling between the SQM
core and the MDM halo governs the overall structure and dynamics of
the star.

This two-fluid formalism has been widely employed in previous studies of
compact stars with dark matter components \cite[e.g.,][]{Sandin2009,Ciarcelluti2011,Leung2011,Li2012,Li2012a,Goldman2013,Xiang2014,Mukhopadhyay2016,Panotopoulos2017,Panotopoulos2018,Ellis2018a,Ellis2018b,Nelson2019,Garani2019,Ivanytskyi2020,Lee2021,Kain2021, Ciancarella2021,Karkevandi2022,Miao2022,Emma2022,Rutherford2023,Diedrichs2023,Ferreira2023,Hippert2023,Giangrandi2023,Rezaei2023,Liu2023,Bramante2024,Shakeri2024,Mariani2024,Hong2024,
Karkevandi2024,Sun2024,Liu2024a,Thakur2024,Barbat2024,Zhen2024,Konstantinou2024,Shawqi2024,Grippa2024}. In
these studies, the dark matter component is treated as a separate
fluid interacting gravitationally with the ordinary matter. The same
approach is used in our model, which treats the SQM and MDM components
as independent fluids, each obeying its own EOS but interacting
through their mutual gravitational field. This approach allows us to
explore the impact of MDM on the mass-radius relation of compact
stars, particularly in the case of XTE J1814-338, which exhibits
unusual mass and radius measurements.

In the following sections, we will explore the mass-radius
relationship and the impact of MDM on SSs, using the EOS
described above. We will also compare our results with recent
observational data to assess the viability of our model in explaining
compact objects such as XTE J1814-338.

\section{Results and Discussion}\label{results}

For the given EOS of SQM,
we compute the structure of SSs admixed with MDM using the two-fluid formalism
\citep{Yang2021b}. This formalism, widely employed in the study of
compact stars containing dark matter, assumes that the SQM and MDM
components interact only through gravity, with no direct interactions
between quarks and mirror quarks. Such a framework allows us to isolate the gravitational influence of the dark matter component and study its impact on the mass-radius relation
of SSs.

\subsection{Mass-Radius Relation of SSs without MDM}

\begin{figure}[tbp]
	\centering
	\includegraphics[width=1.1\linewidth]{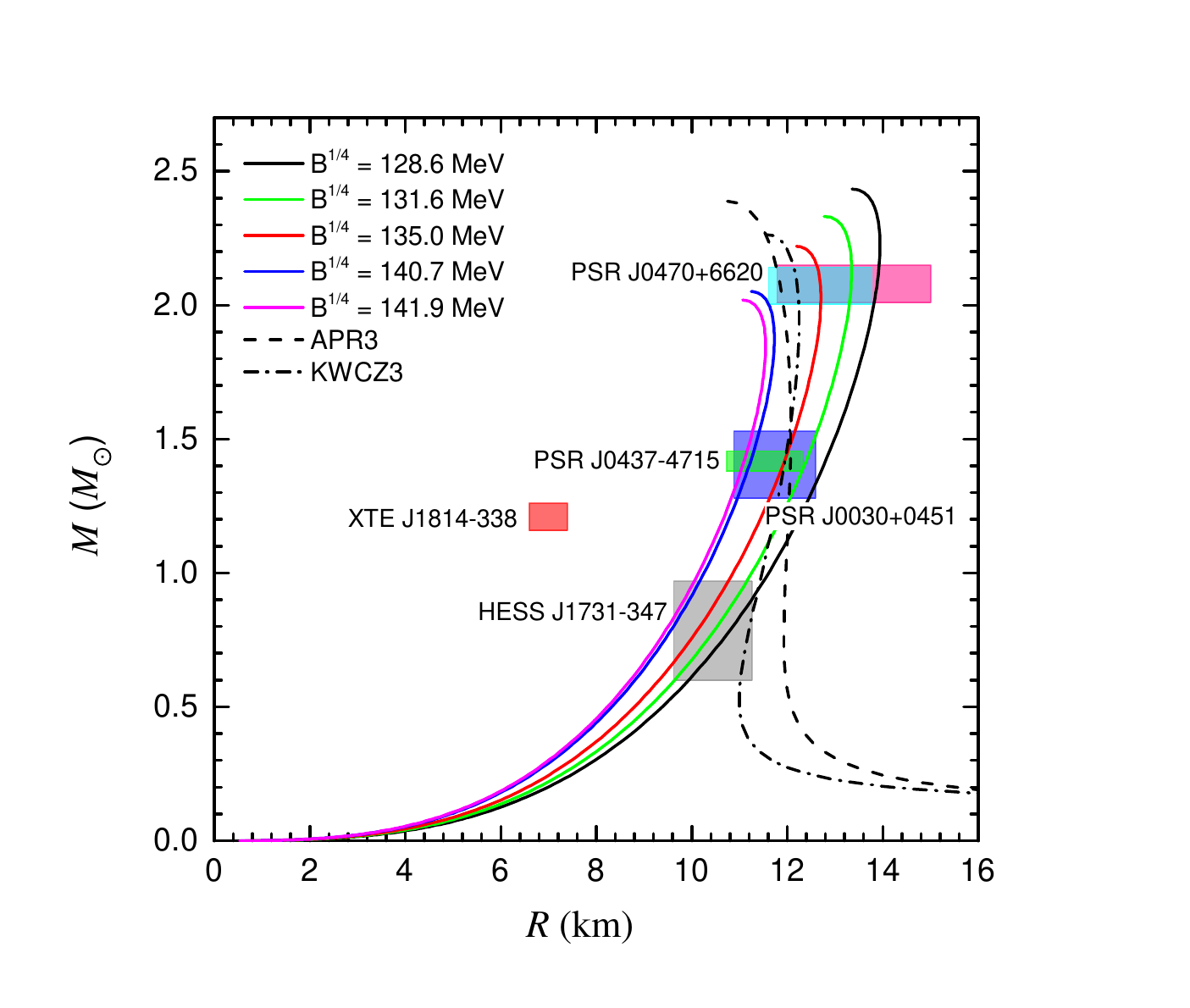}
\caption{The mass-radius relation of SSs without MDM
  (i.e., with a mass fraction of MDM \(f_D = 0\)) for \(\alpha_S =
  0.6\) and different values of \(B^{1/4}\) is shown. The pink and
  cyan regions represent the mass and radius of PSR J0740+6620 as
  presented in Refs. \cite{Cromartie2020,Fonseca2021,Dittmann2024} and Ref. \cite{Salmi2024}, respectively. The blue and green regions show the
  mass and radius of PSR J0030+0451 \citep{Vinciguerra2024} and PSR
  J0437-4715 \citep{Choudhury2024}, respectively. The grey region
  indicates the mass and radius estimates for the central compact
  object within the supernova remnant HESS J1731-347
  \citep{Doroshenko2022}. The red region shows data from the
  observation of XTE J1814-338 \citep{Kini2024}. For comparison, two lines for NSs are also presented: the dashed line corresponds to APR3 EOS \citep{Akmal1998}, and the dash-dotted line corresponds to KWCZ3 EOS (for the case of $C_{\sigma}=14$, $L=40$) \citep{Kubis2023}.}
\label{rm_diff_B} 
\end{figure}

In Fig.\ \ref{rm_diff_B}, we present the mass-radius relation of pure
SSs (i.e., without MDM) for different values of the bag
constant $B^{1/4}$ and a strong interaction coupling constant of
$\alpha_{S}=0.6$. We define the mass fraction of MDM as $f_{D} \equiv
M_{D}/M$, where $M$ is the total mass of the star and $M_{D}$ is the
mass of the MDM component. Figure \ref{rm_diff_B} corresponds to
$f_{D}=0$, meaning no MDM is included.

\begin{figure}[htb]
	\centering
	\includegraphics[width=1.1\linewidth]{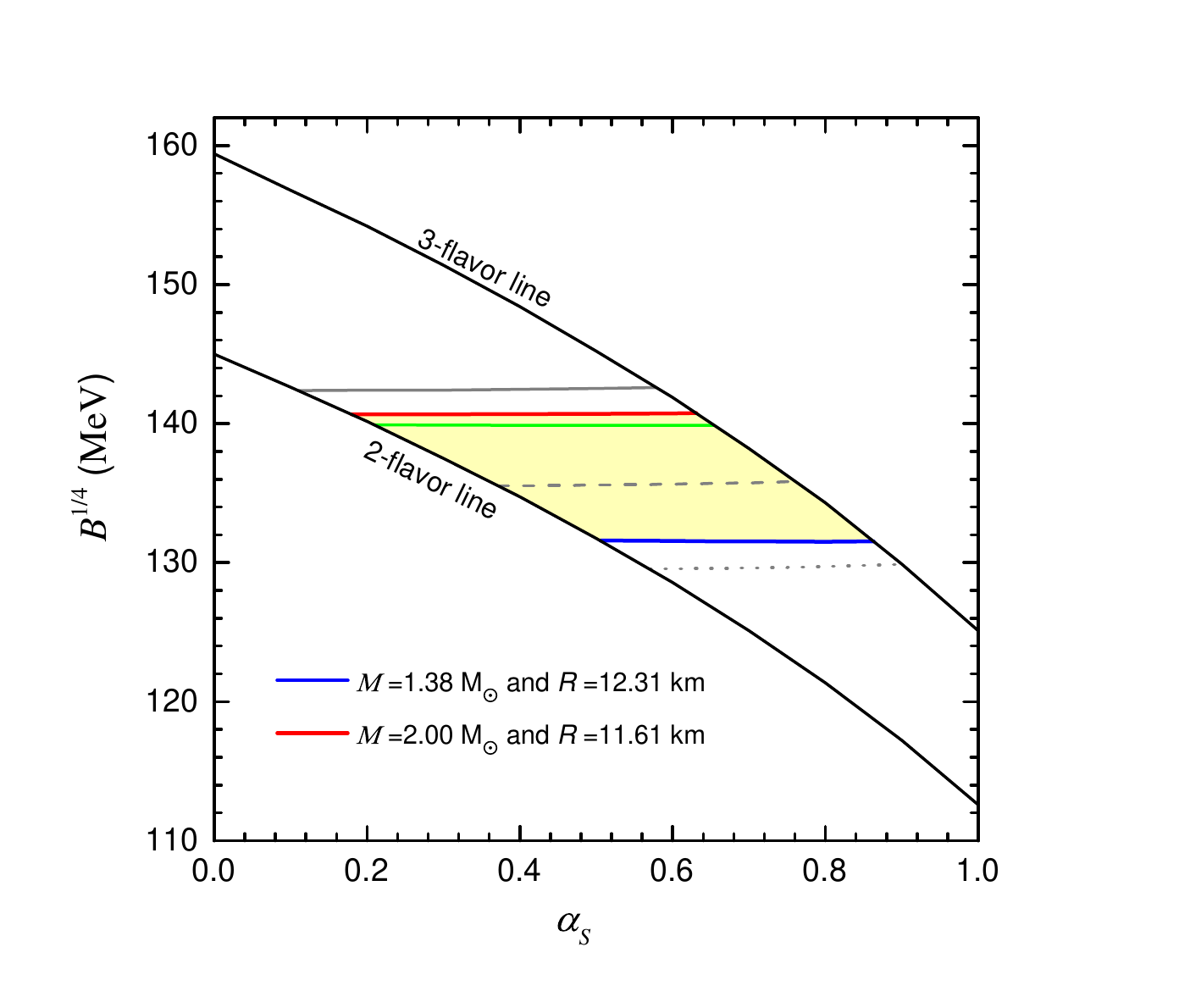}
        \caption{Constraints on \(B^{1/4}\) and \(\alpha_{S}\) for
          SQM. The solid, dashed, and dotted
          grey lines correspond to \(M_{\rm max} = 2.0\, M_{\odot},
          2.2\, M_{\odot},\) and \(2.4\, M_{\odot}\),
          respectively. The green line represents \(\Lambda(1.4) =
          580\) [$\Lambda(1.4)$ is the dimensionless tidal deformability of a 1.4 $M_{\odot}$ pure SS]. All the curves for mass, radius, and tidal
          deformability are calculated for pure SSs
          without MDM, i.e., for \(f_{D} = 0\).}
\label{constraints}
\end{figure}

The range of values for $B^{1/4}$ between 128.6 MeV and 141.9 MeV was
chosen because they satisfy both the "2-flavor line" and "3-flavor
line" constraints, as shown in Fig.\ \ref{constraints}. These
constraints ensure that the SQM remains stable and energetically
favorable in comparison to nuclear matter, which is essential for the
existence of SSs.

From Fig.\ \ref{rm_diff_B}, we observe that the green, red, and blue
curves match most of the current observational data. However, they fail to explain the exotic mass and radius
observation of XTE J1814-338. Specifically, the green line marginally
satisfies the mass and radius constraints for PSR J0437-4715
\citep{Choudhury2024}, while the blue line only marginally satisfies
the observation of PSR J0740+6620 \citep{Salmi2024}. For
$\alpha_{S}=0.6$, the results suggest that the allowed values of
$B^{1/4}$ must fall within the range of 131.6 MeV to 140.7 MeV to
satisfy the majority of observations.

\subsection{Parameter Space and Constraints}

We have further explored the parameter space of the SQM model by
applying four observational constraints
\cite[e.g.,][]{Schaab1997,Weissenborn2011,Zhou2018,Yang2020,Yang2021b,Yang2021a,Yang2023,Yang2024,Pi2025}. These
constraints are essential to ensure that the model can explain the
observed properties of NSs and SSs:

The energy per baryon of SQM must be lower than that of the most
 stable atomic nucleus, $^{56}$Fe, which has an energy per baryon of
 $E/A \sim 930$ MeV \citep{Witten1984,Zhou2018}. This condition
 ensures that SQM is absolutely stable. The parameter region that
 satisfies this constraint is located below the 3-flavor line in
 Fig.\ \ref{constraints}.
   
Non-strange quark matter (i.e., two-flavor quark matter consisting of
only $u$ and $d$ quarks) must have an energy per baryon higher than
$^{56}$Fe, plus a 4 MeV surface correction
\citep{Farhi1984,Zhou2018,Yang2020}. This constraint ensures that
ordinary nuclear matter remains stable and does not dissolve into
quark matter. The region satisfying this constraint lies above the
2-flavor line in Fig.\ \ref{constraints}.

\begin{figure}[tbp]
	\centering
	\includegraphics[width=1.1\linewidth]{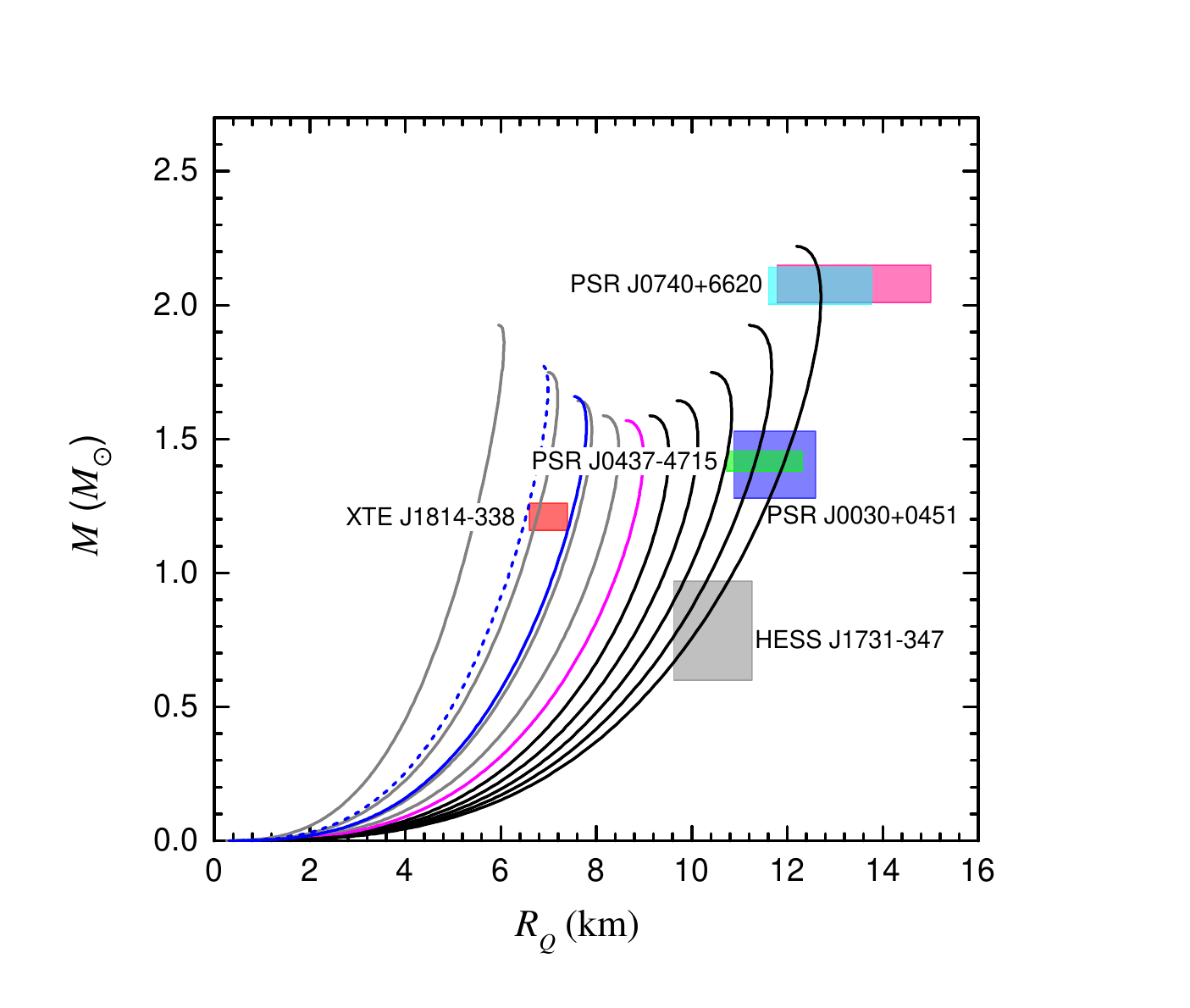}
        \caption{The mass-radius relation (\(R_{Q}\) represents the
          radius of the SQM) of MDM-admixed SSs for \(\alpha_{S} = 0.6\) and \(B^{1/4} = 135\)
          MeV. The magenta line corresponds to \(f_{D} = 50\%\). From
          right to left, the black lines represent \(f_{D} = 0\%\),
          10\%, 20\%, 30\%, and 40\%, while the grey lines represent
          \(f_{D} = 60\%\), 70\%, 80\%, and 90\%. The solid and dashed
          blue lines indicate \(f_{D} = 71.8\%\) and \(f_{D} =
          82.2\%\), respectively. The shaded regions
          are the same as those in Fig.~\ref{rm_diff_B}.}
\label{rm_diff_fd}
\end{figure}

\begin{figure}[tbp]
	\centering
	\includegraphics[width=1.1\linewidth]{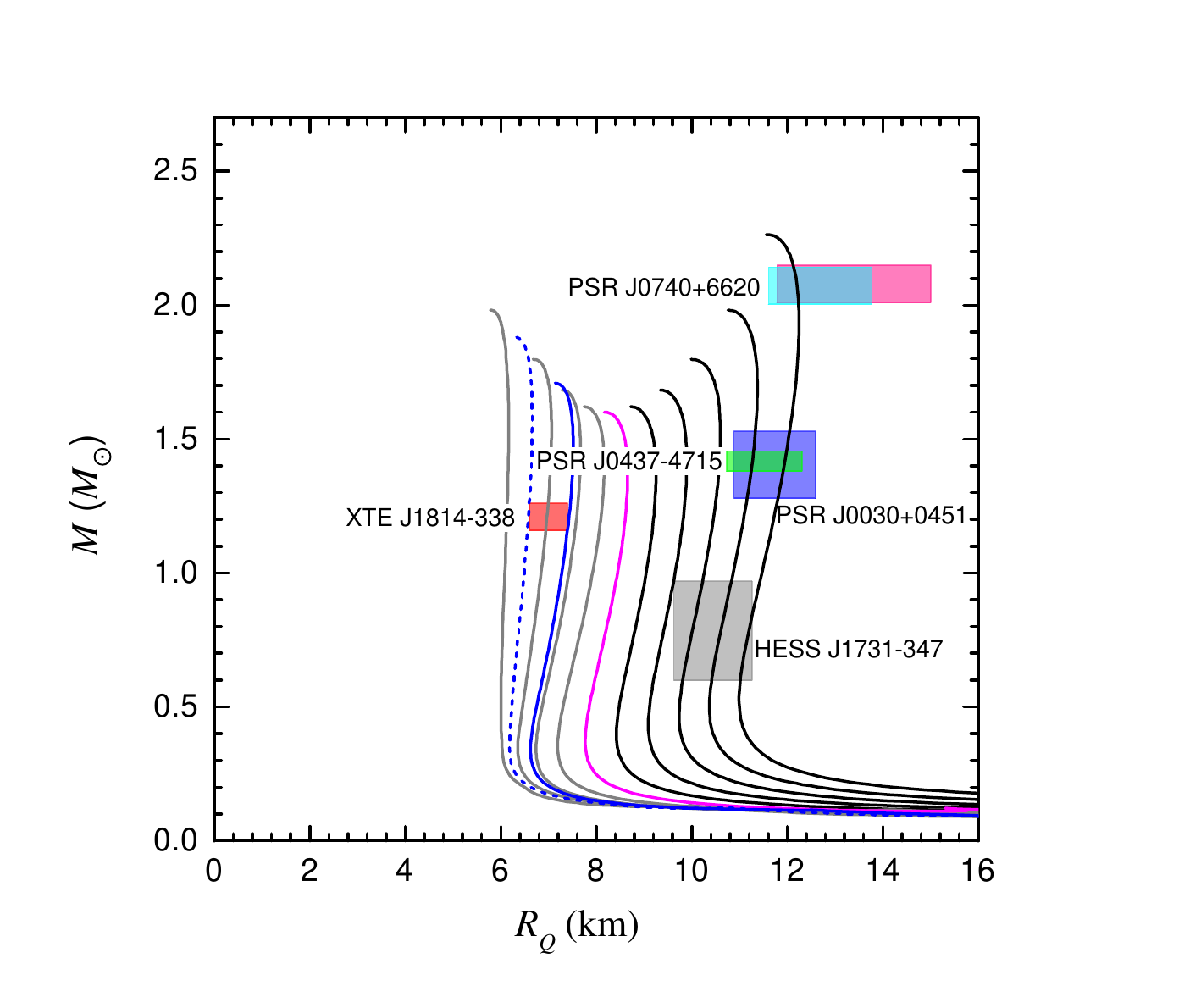}
        \caption{The mass-radius relation (\(R_{Q}\) represents
            the radius of the SQM) of MDM-admixed NSs for the KWCZ3
            EOS. The values of \(f_D\) for the magenta line, and the
            black and grey lines, are the same as those in
            Fig.~\ref{rm_diff_fd}. The solid and dashed blue lines
            indicate \(f_{D} = 72.8\%\) and \(f_{D} = 85.1\%\),
            respectively. The shaded regions are identical to those in
            Fig.~\ref{rm_diff_B}.}
\label{rm_fd_KWCZ}
\end{figure}

The third constraint is derived from the mass and radius measurements
 of PSR J0437-4715 \citep{Choudhury2024}. The region that satisfies
 this constraint is located above the blue line in
 Fig.\ \ref{constraints} (note that the blue line corresponds to $M = 1.38 \, M_{\odot}$ and $R = 12.31$ km, which is the coordinate of the lower-right point of the green rectangle in Fig.~\ref{rm_diff_B}).

The fourth constraint is based on the mass and radius estimates of PSR
J0740+6620 \citep{Salmi2024}. The parameter region satisfying this
constraint is below the red line in Fig.\ \ref{constraints} (the red line corresponds to $M = 2.00 \, M_{\odot}$ and $R = 11.61$ km, which is the coordinate of the lower-left point of the cyan rectangle in Fig.~\ref{rm_diff_B}).

Combining all four constraints, the allowed parameter space is
confined to the yellow-shaded area in Fig.\ \ref{constraints}. Specifically, the parameters of the SQM model are limited to 131.5 MeV $\leq B^{1/4}\leq$ 140.8 MeV (it is slightly different from the case of $\alpha_{S}=0.6$ because different values of  $\alpha_{S}$ are involved here) and 0.18 $\leq \alpha_{S}\leq$ 0.86. It is
important to note that this parameter space is derived under the
assumption that both PSR J0437-4715 and PSR J0740+6620 are pure
SSs without MDM. If MDM were included, the results would
likely differ, potentially broadening the parameter space.

For comparison, the lines for three different values of $M_{\rm max}$ ($M_{\rm max}$ is the maximum mass of SSs without MDM) and $\Lambda(1.4)=580$ [$\Lambda(1.4)$ is the dimensionless tidal deformability of a 1.4 $M_{\odot}$ pure SS] are also shown in Fig.\ \ref{constraints}. The regions below the grey solid, dashed and dotted lines correspond to the parameters that satisfy $M_{\rm max}> 2.0\, M_{\odot}$, $M_{\rm max}> 2.2\, M_{\odot}$, and $M_{\rm max}> 2.4\, M_{\odot}$, respectively. The region above the green line fulfills the $\Lambda(1.4)<580$ constraint from the observation of GW170817 \citep{Abbott2018}.

The constraints outlined above provide a comprehensive framework
  for understanding the properties of SSs under the assumption of pure
  SQM. However, these models do not account for the potential
  influence of additional components, such as mirror dark matter
  (MDM). Incorporating MDM into the equation of state introduces new
  degrees of freedom, significantly altering the mass-radius
  relationship and broadening the parameter space for SSs. To
  investigate these effects, we extend our analysis in the next
  section to include MDM, exploring its impact on the structure and
  observational properties of compact stars.

\subsection{Inclusion of Mirror Dark Matter (MDM)}

While the above constraints are derived for pure SSs, the
inclusion of MDM adds another layer of complexity. As shown in
Fig.\ \ref{rm_diff_fd}, the mass-radius relation ($R_Q$ is the radius of the SQM component, which is the observational radius) of MDM-admixed
SSs ($\alpha_{S}=0.6$, $B^{1/4}=135$ MeV) is strongly
dependent on the mass fraction of MDM, $f_D$. When $f_D \leq 50\%$,
the radius of the MDM component, $R_D$, is smaller than or equal to
 $R_Q$. In this case, the outermost radius of the star corresponds to the SQM
radius. Conversely, when $f_D > 50\%$, the radius of the SQM component
is smaller than the MDM radius, leading to a configuration where the
star is an MSS with an ordinary SQM core.

This two-fluid behavior is crucial for explaining the mass and radius
observations of XTE J1814-338 \citep{Kini2024}. Figure \ref{rm_diff_fd}
shows that the data for XTE J1814-338 can be explained if the star is
an MSS with an ordinary SQM core, where the mass
fraction of MDM falls between $f_D = 71.8\%$ and $f_D = 82.2\%$. This
is consistent with previous results, suggesting that XTE J1814-338
belongs to a class of stars that include both SQM and
MDM.

Figure \ref{rm_fd_KWCZ} illustrates the mass-radius relationship for
MDM-admixed neutron stars, modeled using the KWCZ3 EOS (for the case of $C_{\sigma}=14$, $L=40$) \citep{Kubis2023}. The dependence of the mass-radius curves on the MDM mass
fraction ($f_D$) is evident, with higher $f_D$ values producing more
compact configurations. The data for XTE J1814-338, represented by the
red shaded region, is explained within a specific $f_D$ range of 72.8\%
to 85.1\%. This alignment highlights the potential role of MDM in
shaping the structural properties of compact stars.
 Unlike the MDM-admixed strange stars shown in Fig.\ \ref{rm_diff_fd},
 the neutron star model in Fig.\ \ref{rm_fd_KWCZ} shows subtle shifts in the
 mass-radius curves, particularly at higher $f_D$. The solid and dashed
 blue lines denote the $f_D$ thresholds where the radius of the neutron
 star’s baryonic component becomes dominated by the MDM envelope,
 signifying a transition to a two-layer structure.
The inclusion of MDM alters the tidal deformability and compactness of
neutron stars, making these models distinguishable from traditional
neutron stars. For XTE J1814-338, the high $f_D$ values required to
explain its observations suggest a substantial dark matter admixture,
underscoring the critical role of multi-messenger astrophysical
observations.

\section{Considerations and Future Perspectives}\label{sec:testing}

The predictions of our models of SSs admixed with MDM can be thoroughly tested through multi-messenger
astronomy, which combines gravitational wave, X-ray, and thermal
emission observations. These observations offer unique opportunities
to validate the presence of exotic matter compositions in compact
stars, such as XTE J1814-338. In the following, we explore how
gravitational wave signatures, precise mass-radius measurements from
X-ray timing, and thermal emission studies can be used to test and
refine these predictions.

Gravitational wave detections provide a powerful method to probe the
internal structure of compact stars. Events such as GW170817 \citep{Abbott2017} have
constrained tidal deformability, setting the stage for testing our
model. The SS+MDM configuration is predicted to exhibit distinctive
tidal deformability signatures in future observations of binary star
mergers. These signatures, observable with advanced detectors like
LIGO, Virgo, Kagra, and next-generation observatories (e.g., Einstein
Telescope and Cosmic Explorer), could serve as decisive evidence
supporting the model.

X-ray timing observations from missions like NICER are crucial for
assessing the SS+MDM model. Precise mass and radius measurements, such
as those for PSR J0740+6620 and XTE J1814-338, allow direct
comparisons with our predictions of compactness and the effects of MDM
admixture. Incorporating improved measurements and reducing
observational uncertainties will help refine parameter estimates such
as the bag constant $B$ and MDM fraction $f_D$.

The presence of an MDM halo is expected to influence the thermal
evolution and cooling behavior of compact stars. Deviations from
standard cooling curves could provide additional observational
signatures pointing to MDM. This avenue offers complementary insights
and underscores the importance of thermal emission studies alongside
mass-radius observations.

While our model provides a compelling explanation for XTE
  J1814-338, alternative frameworks—such as boson stars \cite{Karkevandi2024,Pitz2024} or fermionic
  dark matter models \cite{Liu2024a,Jockel2024}—also predict compact objects  with unusual
  mass-radius properties.  Boson stars may involve new scalar
  particles beyond the Standard Model, which could introduce novel
  physics but also pose challenges for physical
  interpretation. Fermionic dark matter stars rely on self-interacting
  fermions to achieve compactness, a feature that invites further
  exploration. In contrast, the SS+MDM framework builds upon known
  physical principles and extends the strange star hypothesis in a
  natural way. However, like all models, the SS+MDM framework also
  requires further validation and refinement. Future observations,
  including measurements of tidal deformability, cooling curves, and
  gravitational wave signatures, will be critical for testing and
  distinguishing between these models.

Ongoing and future missions, including eXTP \citep{Zhang2019,Watts2019}, STROBE-X \citep{Ray2019} and ATHENA \citep{Majczyna2020},
promise enhanced observational precision. These advancements will
enable more accurate tests of the SS+MDM framework and help constrain
the parameter space of exotic compact objects. Simultaneously,
theoretical efforts should explore alternative dark matter candidates
and potential direct interactions between SQM and MDM to refine
existing models.

Ultimately, multi-messenger astronomy has the potential to provide
valuable insights for evaluating the predictions of the SS+MDM model,
distinguishing it from competing theories, and deepening our
understanding of compact star structures and the role of dark matter
in astrophysics.

\section{Summary}\label{Summary}

In this paper, we have investigated the structure and properties of
SSs admixed with MDM and their implications for recent NS
observations. Our study demonstrates that this novel approach,
incorporating MDM into the EOS of SQM, provides a viable explanation
for the observed mass and radius measurements of several key
astrophysical objects.

To assess the robustness of our SS+MDM framework, we conducted a
sensitivity analysis of key parameters, such as the bag constant B and
the MDM mass fraction $f_D$. Variations in $B$ within the allowed
range (131.5 MeV to 140.8 MeV) were found to significantly affect the
predicted radii of SSs, with lower values of $B$ leading to more
compact configurations. Similarly, changes in the strong interaction
coupling constant $\alpha_S$ introduced shifts in the mass-radius
curves, emphasizing the importance of constraining these parameters
through independent observations. For the MDM component, the mass
fraction $f_D$ showed the most pronounced influence on the model’s
ability to reproduce the observed properties of XTE J1814-338. Stars
with $f_D$ outside the range of 71.8\% to 82.2\% were unable to match
the measured compactness of XTE J1814-338.

Building on this analysis, we first applied our framework to explain
the mass and radius observations of PSR J0740+6620, PSR J0030+0451,
PSR J0437-4715, and the central compact object within the supernova
remnant HESS J1731-347. These objects were modeled as pure SSs without
requiring additional dark matter components. The results demonstrate
that standard SSs can  account for the majority of the observational data,
supporting the hypothesis that SQM may represent the true
ground state of baryonic matter. This consistency with previous
studies reinforces the plausibility of SSs as an explanation for these
NS observations.

Horvath et al. \cite{Horvath2023} examined XMMU J173203.3-344518 in the remnant
HESS J1731-347, proposing it as an SS due to its low mass and
small radius, which challenge traditional NS models. Their
analysis, based on quark matter equations of state and
color-flavor-locked phases, supports the idea that SSs
could explain both low-mass and heavier compact objects. This aligns
with our findings for XTE J1814-338, strengthening the case for
SQM in exotic compact stars. Our model, incorporating
MDM, complements their work by offering a
mechanism to explain unusual compact star properties, expanding the
scope of SS candidates.

Recently, Giangrandi et al. \cite{Giangrandi2023} investigated the effects of
self-interacting bosonic dark matter on NS properties, using
a two-fluid model to describe the dark matter and baryonic matter
components. Their findings reveal that the presence of dark matter
significantly alters the mass, radius, and tidal deformability of
NSs, depending on the distribution of dark matter in the
core or halo. These results align with our hypothesis of MDM admixed with SSs, further supporting the idea that
dark matter may play a crucial role in shaping the properties of
compact stellar objects.

XTE J1814-338, with its unique mass and radius, presents a challenge
for standard NS models. Our study suggests this source can
be explained as an MSS, with an SQM core and an MDM envelope, consistent with an MDM mass
fraction of $f_{D} = 71.8\%$ to $f_{D} = 82.2\%$. This provides a compelling case for the
presence of MDM in compact stars and positions XTE J1814-338 as a
prime candidate for further dark matter investigations in
astrophysical environments.

XTE J1814-338 can also be explained by alternative models, such as the
boson star model proposed by Pitz and Schaffner-Bielich \cite{Pitz2024}, which
describes a core of ordinary nuclear matter, or by compact stars
admixed with self-interacting fermionic dark matter, as shown by Liu et al. \cite{Liu2024a}.
While these models provide alternative explanations,
the success of our MDM model highlights the viability and potential
necessity of dark matter, particularly in the form of MDM, in
explaining such exotic observations.

For the explaination of the observation of XTE J1814-338, large values of $f_{D}$ (lager than 70\%) should be employed, which could not be obtained
from normal accretion processes \cite{Karkevandi2022}. However, it is reachable through other mechanisms listed in ref. \cite{Karkevandi2022}, e.g.,
it could be realized if a normal SS with a smaller mass mergers with a pure MSS with a larger mass.

As the field progresses, developing methods to differentiate between
various dark matter candidates, such as boson stars, fermionic dark
matter, and MDM, will be essential. Upcoming
observations from advanced gravitational wave detectors such as aLIGO, aVirgo, Kagra, the Einstein Telescope (ET), and the Cosmic Explorer (CE); and the X-ray
missons such as eXTP \citep{Zhang2019,Watts2019}, STROBE-X \citep{Ray2019} and ATHENA \citep{Majczyna2020}, are likely to provide the sensitivity required to detect
subtle differences between these models, especially in the context of
multi-messenger astronomy.

\begin{acknowledgements} We thank Sebastian Kubis for kindly providing the KWCZ EOS table. This work is supported by the National Key R\&D
Program of China (Grant No.\ 2021YFA0718504) and the Scientific
Research Program of the National Natural Science Foundation of China
(NSFC, Grant No.\ 12033001).
\end{acknowledgements}

\bibliographystyle{apsrev4-1} 
\bibliography{references}

\end{document}